\documentclass{vgtc}                          

\ifpdf
  \pdfoutput=1\relax                   
  \usepackage{graphicx}                
  \DeclareGraphicsExtensions{.pdf,.png,.jpg,.jpeg} 
\else
  \usepackage{graphicx}                
  \DeclareGraphicsExtensions{.eps}     
\fi%

\graphicspath{{figures/}{pictures/}{images/}{./}} 

\usepackage{microtype}                 
\PassOptionsToPackage{warn}{textcomp}  
\usepackage{textcomp}                  
\usepackage{mathptmx}                  
\usepackage{times}                     
\usepackage{cite}                      
\usepackage{tabu}                      
\usepackage{booktabs}                  

\onlineid{0}

\vgtccategory{Research}

\vgtcinsertpkg

\title{Auditory Feedback to Make Walking in Virtual Reality More Accessible}

\author{M. Rasel Mahmud\thanks{e-mail: m.raselmahmud1@gmail.com}\\  %
     \parbox{1.4in}{\scriptsize \centering Computer Science \\ The University of Texas at San Antonio} %
\and Michael Stewart\thanks{e-mail: michael.stewart@utsa.edu }\\ %
     \parbox{1.4in}{\scriptsize \centering Kinesiology \\ The University of Texas at San Antonio} %
\and Alberto Cordova\thanks{e-mail: Alberto.Cordova@utsa.edu}\\ %
     \parbox{1.4in}{\scriptsize \centering Kinesiology \\ The University of Texas at San Antonio} %
\and John Quarles\thanks{e-mail: John.Quarles@utsa.edu}\\ %
     \parbox{1.4in}{\scriptsize \centering Computer Science \\ The University of Texas at San Antonio}}

\abstract{
The objective of this study is to investigate the impact of several auditory feedback modalities on gait (i.e., walking patterns) in virtual reality (VR). Prior research has substantiated gait disturbances in VR users as one of the primary obstacles to VR usability. However, minimal research has been done to mitigate this issue. We recruited 39 participants (with mobility impairments: 18, without mobility impairments: 21) who completed timed walking tasks in a real-world environment and the same tasks in a VR environment with various types of auditory feedback. Within-subject results showed that each auditory condition significantly improved gait performance while in VR  (\textit{p} $<$ .001) compared to the no auditory condition in VR for both groups of participants with and without mobility impairments. Moreover, spatial audio improved gait performance significantly (\textit{p} $<$ .001) compared to other auditory conditions for both groups of participants. This research could help to make walking in VR more accessible for people with and without mobility impairments.
} 


\CCScatlist{
  \CCScatTwelve{Human-centered computing}{Human computer interaction (HCI)}{interaction paradigms}{Virtual reality};
  \CCScatTwelve{Human-centered computing}{Human computer interaction (HCI)}{HCI design and evaluation methods}{User studies}
}

\begin{document}


\firstsection{Introduction}

\maketitle


Virtual reality (VR) technology has a wide variety of applications (e.g., education, physical fitness, rehabilitation, entertainment). However,  VR causes gait (i.e., walking patterns) disturbance in most users, which limits the usability and benefits of VR \cite{agrawal2009disorders,ferdous2018investigating,guo2013effects}. This problem is especially severe for persons with mobility impairments (MI) as these populations have functional gait disorders, and further gait disturbance in VR makes it increasingly difficult for them to use VR technologies. For example, persons with mobility impairments may find it very difficult to perform various locomotor movements in VR without the risk of falling or injury. However, minimal research has been conducted to mitigate these challenges. 

Outside of VR research, the field of assistive technologies has shown that some multimodal feedback techniques \cite{franco2012ibalance,sienko2017role} can improve gait and balance and support individuals with MI during daily activities. For example, assistive technology based on vibrotactile \cite{mahmud2022vibrotactile} and visual feedback has been applied in studies aimed at improving balance and gait for persons with disabilities \cite{velazquez2010wearable,thikey2011need,vcakrt2010exercise,sutbeyaz2007mirror}. Similarly, auditory feedback in a non-VR environment has also improved gait in some prior studies. For example, Baram et al. \cite{baram2007auditory} reported that walking velocity and stride length improved significantly in a non-VR environment using auditory feedback compared to baseline for participants with Multiple Sclerosis (MS). However, the application of auditory feedback on gait and balance performance in VR has not been thoroughly explored. 

To investigate solutions to the gait disturbance issue, we conducted empirical studies applying various auditory feedback techniques (e.g., spatial, static rest frame, and rhythmic audio) in VR for participants with and without MI. Participants performed a timed walking task using a pressure-sensitive walkway for quantitative gait analysis -  GAITRite. In our study, all auditory conditions improved gait parameters significantly whereas spatial audio outperformed others. The purpose of this study was to make immersive VR more accessible using auditory feedback and analyzing its influence on gait performance while in VR. However, we did not measure post-study effects on gait performance.

\section{Background and Related Work}

\subsection{Gait Disturbance in VR}
VR has been shown to induce instability and gait disturbance in prior studies. Research published in  2001 reported that balance in VR was impaired \cite{takahashi2001change}. Individuals using HMDs can lose stability due to end-to-end latency and illusory impressions of body movement generated by VR because HMDs obstruct visual feedback from the real world \cite{soltani2020influence, martinez2018analysing}. Prolonged engagement in VR also resulted in postural instability\cite{murata2004effects}. These postural instabilities caused by VR can induce gait instability while walking in a virtual environment (VE) \cite{hollman2007does}. Riem et al. \cite{riem2020effect} also reported significant disturbance of step length (\textit{p} $<$ .05) in VR compared to baseline. Other studies have also explored the impact of imbalance on gait disturbance \cite{sondell2005altered}. Horsak et al. recruited 21 participants (male: 9, female:12, age: 37.62 ± 8.55 years) to see if walking in an HMD-based VE has a significant impact on gait\cite{horsak2021overground}. Walking speed was reduced by 7.3\% in HMD-based VE in their study. Canessa et al. also investigated the difference between real-world walking and immersive VR walking using an HMD\cite{canessa2019comparing}. They reported that walking velocity decreased significantly (\textit{p} $<$ .05) in immersive VE compared to real-world walking.  Also, most prior studies concentrated on participants without MI \cite{lott2003effect,epure2014effect,robert2016effect,horlings2009influence,samaraweera2015applying} For example, Martelli et al.\cite{martelli2019gait} investigated whether gaits change during overground walking in a VE while using a VR HMD with continuous multidirectional visual field perturbations. In four different settings, 12 healthy young adults walked for six minutes on a pathway. Reduced stride length, greater stride width, and higher stride variability were observed when the visual field was perturbed. 
However, there have been very few attempts in the past to address gait disturbance in VR. This inspired us to investigate the gait disturbance issue to make walking in VR more accessible.

\subsection{Gait Improvement After VR Intervention}
Although the focus of our research is VR accessibility and gait improvement while in VR, it is important to review how VR rehabilitation applications have previously facilitated balance and gait improvement that persists after the VR experience is over \cite{de2016effect,meldrum2012effectiveness,park2015effects,cho2016treadmill,duque2013effects, bergeron2015use}.
For example, Walker et al. \cite{walker2010virtual} recruited seven post-stroke patients with MI to investigate the improvement in walking and balance abilities using a low-cost VR system. They designed the VE to provide participants the sensation of walking along a  city street, which was displayed via a television screen in front of a treadmill. They collected postural feedback via a head-mounted position sensor. An overhead suspension harness supported all participants. Six participants (mean age 53.5y, range 49–62y) completed the study. Results suggested significant improvement (\textit{p} $<$ .05) in post-study balance, walking speed, and gait functionality. Berg Balance Scale (BBS) score improved by 10\%, walking speed improved by 38\%, and Functional Gait Assessment (FGA) score increased by 30\% compared to baseline in their study. However, the majority of the prior work in VR gait rehabilitation did not use HMDs. We used HMDs to render the VEs in our study.

\subsection{The Effects of HMDs on Persons With Gait Disturbance}
Winter et al. \cite{winter2021immersive} recruited 36 (Male: 10, Female: 26) participants without MI and 14 participants with MI (MS: 10, Stroke: 4) to investigate the effect of an immersive, semi-immersive, and no VR environment on gait during treadmill training. First, participants completed the treadmill training without VR. Then, they experienced a virtual walking path displayed via a monitor in the semi-immersive VR condition. They experienced the same VR scenario via HMD in the immersive VR condition. Experimental results showed that immersive VR during gait rehabilitation increased walking speed more significantly (\textit{p} $<$ .001) than semi-immersive and no VR conditions for participants with and without MI. Participants did not experience cybersickness or a significant increase in heart rate after the VR conditions. 

Janeh et al. recruited 15 male patients with Parkinson's disease to investigate a VR-based gait manipulation approach aimed at achieving gait symmetry by adjusting step length \cite{janeh2019gait}. They compared natural gait with walking circumstances during VR-based gait manipulation activities utilizing visual or proprioceptive signals. VR manipulation activities enhanced step width and swing time as compared to natural gait. Janeh et al. also reported VR as a promising and potentially beneficial tool for improving the gait of persons with neurological disorders after VR experience. They stressed the significance of using virtual walking approaches in rehabilitation\cite{janeh2021review}.

Also, Guo et al. \cite{guo2015mobility}
investigated the effect of VEs on gait for both participants with and without MI. They reported that MI participants responded differently in terms of walking velocity, step length, and stride length. However, there was no significant difference for other gait parameters between participants with and without MI. 

Ferdous et al. \cite{ferdous2018investigating} investigated the effect of HMDs and visual components on postural stability in VR for participants with MS. However, they did not investigate the effect on gait, which is the case of most prior studies in immersive VR with HMDs. As a result, the impacts of immersive VR with HMDs on gait parameters have not received enough attention, prompting us to look into the effect on gait in VEs with HMDs for people with and without disabilities.



\subsection{Non-VR Assistive Technology: Auditory Feedback for Gait and Balance Improvement in Real World}
Prior research in non-VR environments found that auditory feedback can greatly improve postural control in the real world, although it is considered less effective than visual feedback techniques \cite{gandemer2016sound}.
Auditory feedback based on the user's lateral trunk lean helped to maintain postural stability \cite{chiari2005audio}. \textit{Spatial audio} - audio that is localized in 3D by the user - was effective in preserving postural stability \cite{stevens2016auditory,gandemer2017spatial}. \textit{Static rest frame audio} - white noise that is uniform and continuous - was found to reduce postural instability in older adults\cite{ross2016auditory,cornwell2020walking}. People having mobility impairments (e.g., people with multiple sclerosis, Parkinson's) and the elderly improved gait using \textit{rhythmic audio} (hearing a consistent beat) \cite{ghai2018effect}. Maculewicz et al. also investigated different rhythmic auditory feedback patterns\cite{maculewicz2015effects}. They reported a significant increase in walking speed (\textit{p} $<$ .001) with the rhythmic auditory feedback compared to the without auditory feedback condition.  However, spatial audio was employed more frequently than other auditory approaches since it is claimed to be more natural and realistic \cite{chong2020audio, pinkl2020spatialized}. However, these studies investigated the auditory feedback in non-VR settings whereas we investigated the effect of auditory feedback on gait improvement in VR settings.

\subsection{Auditory Feedback in VR and the Effect on Gait}
Limited research has been done to investigate the effect of auditory feedback on gait and balance in VR. Mahmud et al. \cite{mahmud2022auditory} investigated different auditory feedback to mitigate imbalance issue in VR. They observed that all auditory feedback improved balance significantly for both participants with and without MS while spatial audio outperformed others. Gandemer et al. studied persons with low vision and found that spatial audio in an immersive VE enhanced gait and balance \cite{gandemer2017spatial}. In most cases, spatial audio was favored for usage in VR because it gave more immersion \cite{mahmud2022auditory, wenzel2017perception, naef2002spatialized}. 
However, the effect of various auditory feedback on VR walking has been minimally studied \cite{nilsson2018natural}. This inspired us to explore the impact of spatial, static rest frame, and rhythmic auditory feedback on gait in VR with a specific focus on persons with gait-related disabilities.

\section{Methods}

\subsection{Hypotheses}
The impact of three auditory techniques (spatial, static rest frame, and rhythmic) on gait parameters in a VR environment was explored in this study. Spatial\cite{mahmud2022auditory, gandemer2017spatial}, Static rest frame \cite{ross2016auditory,cornwell2020walking}, rhythmic \cite{ghai2018effect} auditory feedback types were found to be effective in previous literature in VR and non-VR settings, which motivated us to choose these auditory feedback conditions. These hypotheses were largely motivated by the literature on gait disturbances in VR and non-VR auditory techniques to improve gait in the real world (see Background and Related Work).

H1: Gait disturbances will happen in VR baseline without auditory techniques as compared to the non-VR baseline without auditory techniques.

H2: Three VR-based auditory techniques (spatial, static rest frame, and rhythmic) will improve gait parameters more than the no-audio in VR condition.  

H3: Spatial audio technique will improve gait parameters more than static rest frame and rhythmic audio techniques.

H4: Gait improvement (e.g., velocity) might be more apparent in participants with MI than participants without MI.

\subsection{Participants, Selection Criteria, and Screening Process}
We recruited 39 participants (Male: 9, Female: 30) from various multiple sclerosis support groups and the local community using a matched-comparison group design to investigate gait improvement using auditory feedback in VR. Of these, eighteen participants (Male: 5, Female: 13) had MI due to multiple sclerosis. 52.4\% of the participants with MI identified as White, 28.6\% identified as Hispanic, and 23.8\% identified as African American. In addition, we recruited a group of twenty-one participants without MI (Male: 4, Female: 17). 19\% of the participants without MI were White, 52.4\% were Hispanic, 28.6\% were African American, 9.5\% were American Indian, and 9.5\% were Asian. Both participant groups were statistically comparable in age, height, and weight. Table 1 displays participants' mean (SD) age, height, weight, and gender characteristics for both participant groups. We excluded participants with cognitive impairments, severely low vision, cardiovascular or respiratory conditions, or the inability to walk without assistance. The main challenge for this study was to recruit participants with MS. 

\begin{table}[ht!]
\caption{Descriptive statistics for participants}
    \label{tab:my_label}

\begin{tabular}{|l|l|l|l|l|l|}
\hline
\begin{tabular}[c]{@{}l@{}}Group\\ Name\end{tabular}            & \begin{tabular}[c]{@{}l@{}}No. \\ of \\ Male\end{tabular} & \begin{tabular}[c]{@{}l@{}}No.\\ of\\ Female\end{tabular} & \begin{tabular}[c]{@{}l@{}}Age \\ (Years)\\ Mean \\ (SD)\end{tabular} & \begin{tabular}[c]{@{}l@{}}Height \\ (cm)\\ Mean \\ (SD)\end{tabular} & \begin{tabular}[c]{@{}l@{}}Weight \\ (Kg)\\ Mean \\ (SD)\end{tabular} \\ \hline
\begin{tabular}[c]{@{}l@{}} Participants \\ with MI\end{tabular}     & 5                                                         & 13                                                        & \begin{tabular}[c]{@{}l@{}}44.8 \\ (13.2)\end{tabular}                & \begin{tabular}[c]{@{}l@{}}163.32 \\ (12.64)\end{tabular}                    & \begin{tabular}[c]{@{}l@{}}81.87 \\ (23.63)\end{tabular}              \\ \hline
\begin{tabular}[c]{@{}l@{}} Participants \\ without MI\end{tabular} & 4                                                         & 17                                                        & \begin{tabular}[c]{@{}l@{}}43.2 \\ (12.6)\end{tabular}                & \begin{tabular}[c]{@{}l@{}}164.33 \\ (12.7)\end{tabular}                     & \begin{tabular}[c]{@{}l@{}}85.25 \\ (17.96)\end{tabular}              \\ \hline
\end{tabular}
\end{table}

\textbf{\textit{Screening Process:}}
Participants were recruited through telephone calls, email lists, and flyers. Pre-screening was conducted over the telephone to determine participants' eligibility. We inquired about general demographic information, health, and medical-related history to address participant inclusion or exclusion from the study. For example, we confirmed the individual’s ability to visit the on-campus lab and participate through the duration of the study. We also assessed participants' history of MI and their ability to walk without assistance. We minimized participant characteristic imbalances and ensured age, height, and weight were proportionally similar between both participant groups.

\subsection{System Description}
The following equipment was used in the study for participants' safety and data collection.

\textbf{\textit{Computers, VR Equipment, and Software:}}
The VEs were developed using Unity3D software. We used an HTC Vive wireless HMD which has a pixel resolution of 2160 x 1200, 90 Hz refresh rate, and a 110-degree field of view. We used the integrated HMD headphones to apply the auditory feedback techniques. We used a computer to render the VE with specifications including a Windows 10 operating system, an Intel Core i7 Processor (4.20 GHz), 32GB DDR3 RAM, and an NVIDIA GeForce RTX 2080 graphics card.

\textbf{\textit{Safety Equipment:}}
We used a Kaye Products Inc. suspension walking system which consisted of a body harness, thigh cuffs, and suspension walker for the safety of the participants during the study.

\textbf{\textit{Gait Analysis:}}
A GAITRite walkway system was used to collect participants' gait parameters. The GAITRite walkway system is a portable 12 ft. pressure sensor pad capable of providing spatial and temporal gait parameters of participants during a walking task.

\textbf{\textit{Environment:}}
The study was conducted in a controlled environment ($>$ 600sq ft.). We conducted each study with only the participant and researcher in the room in order to minimize any ambient noises or other disturbances. Fig. 1 shows the comparison between the real-world environment and the virtual environment for the timed walking task.

\begin{figure}[h!]
    \centering
    \includegraphics[width=0.20\textwidth, height=8.79cm, angle=270]{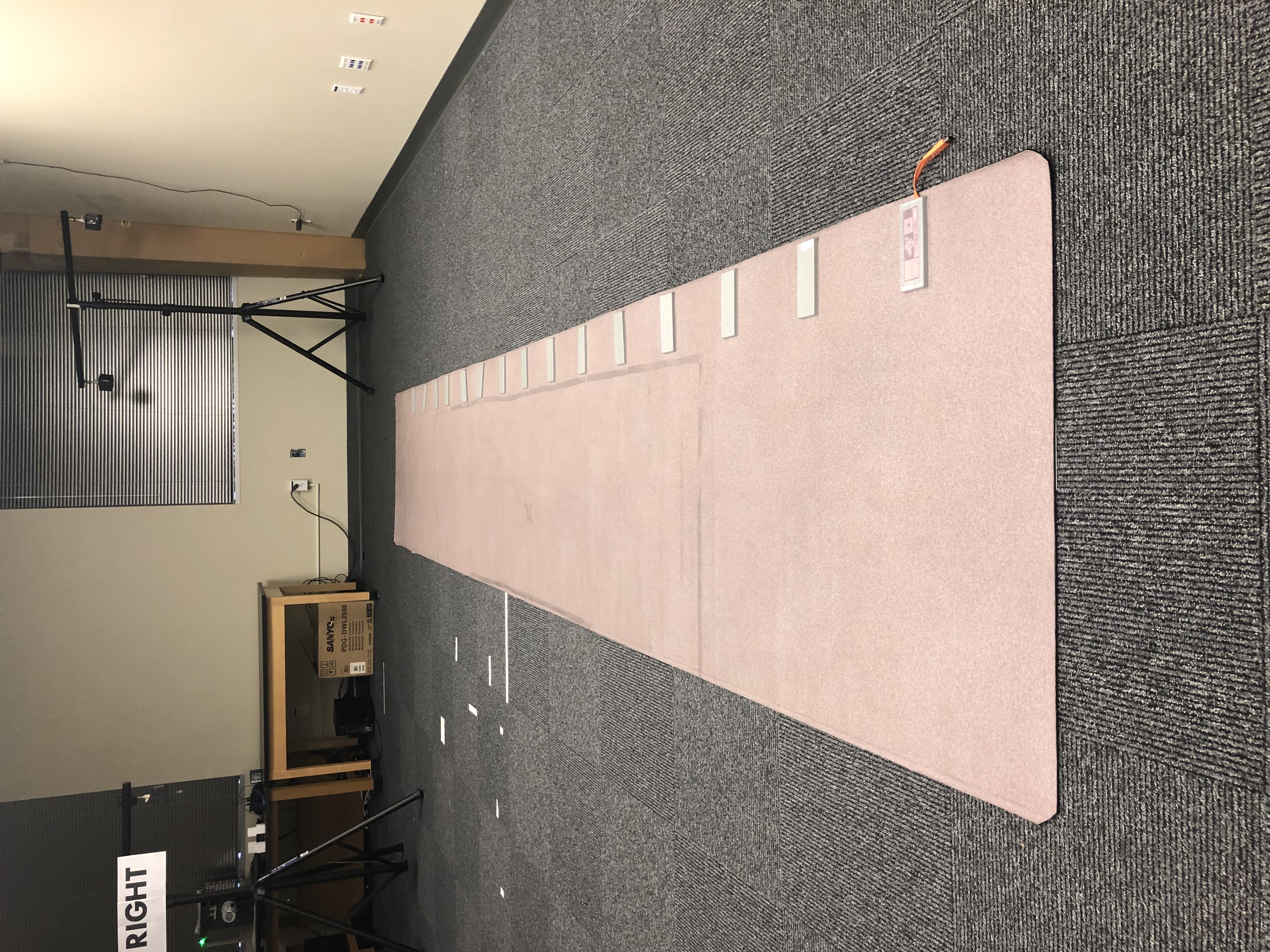}
  \includegraphics[width=0.495\textwidth,height=4cm]{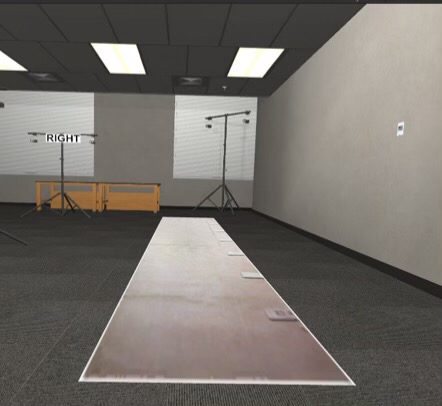}
  \caption{ Comparison between real environment (top) and virtual environment (bottom) for timed walking task}
\end{figure}

\subsection{Study Conditions}
We applied three VR-based auditory feedback techniques and one condition with no audio in order to observe the effect of these feedback techniques on an individual's gait performance. We played these feedback conditions at the start of the tasks through the HMD's integrated headphones. We employed white noise for auditory feedback because it had been found to improve gait and balance performance due to stochastic resonance phenomena \cite{helps2014different}. In prior research in the real world, auditory white noise was also found to be beneficial in minimizing postural instability \cite{cornwell2020walking, harry2005balancing, sacco2018effects,zhou2021effects,ross2015auditory}. The study conditions were:

\subsubsection{Non-VR Baseline}
We measured the baseline data while participants performed the timed walking task using the GAITRite system without any auditory feedback.

\subsubsection{VR Baseline}
We performed the same timed walking task in VR but with no auditory feedback to establish a VR baseline measurement for participants. Participants performed this condition while wearing the HMD and integrated headphones.

\subsubsection{Auditory VR Feedback}
We applied the following three auditory conditions in VR.

\textbf{\textit{Spatial Audio:}}
This was 3D auditory feedback in relation to the participant's physical position in the lab.  This was simulated spatial audio (rather than recorded ambisonic audio). In particular, we played spatialized white noise from Unity3D such that when the user rotated their head, the noise played at varying levels in each ear to imitate a stationary sound source. We used Google resonance audio SDK to implement this because the plugin utilizes head-related transfer functions (HRTFs) to simulate 3D sound more accurately than unity's default \cite{WinNT1}. The 3D audio source in the VE had X, Y, and Z coordinates of 0, 0, and 0, respectively.

\textbf{\textit{Static Rest Frame Audio:}}
This auditory feedback was continuous white noise that was not relative to the participants positioning in the lab. Participants heard the noise at the same intensity in both ears all the time for this condition which is similar to a "mono" sample with no panning. In previous non-VR research, this strategy was also shown to enhance adult participants' balance. \cite{ross2016auditory, cornwell2020walking}.

\textbf{\textit{Rhythmic Audio:}}
The white noise was similar to the static rest frame, but it was a white noise clip at every one-second interval. The length of the rhythmic audio clip was also one-second. Previous research revealed that hearing a constant rhythm may enhance balance and walking in persons with neurological disorders and the adults in non-VR settings \cite{ghai2018effect}.

\subsection{Auditory Feedback Design}
In Unity3D, we connected the audio to sound sources and modified them to fit our research needs. We had a different Unity scene for each auditory feedback circumstance. When the participant was ready, we started each condition with the relevant scene to start the auditory feedback. We performed the scenes for all participants in counterbalanced order, which assigned the auditory feedback in different orders for different participants. We used counterbalancing as it reduces carryover and fatigue effects\cite{WinNT5,WinNT6,WinNT4}. The audio was delivered over the wireless HMD's embedded headphones at the start of the task. The loudness of the audio was adjusted to the participant's satisfaction.

\subsection{Study Procedure}
The flowchart in Fig. 2 represents the whole study procedure. 

\begin{figure}[ht!]
    \centering
  \includegraphics[width=0.47\textwidth,height=10cm]{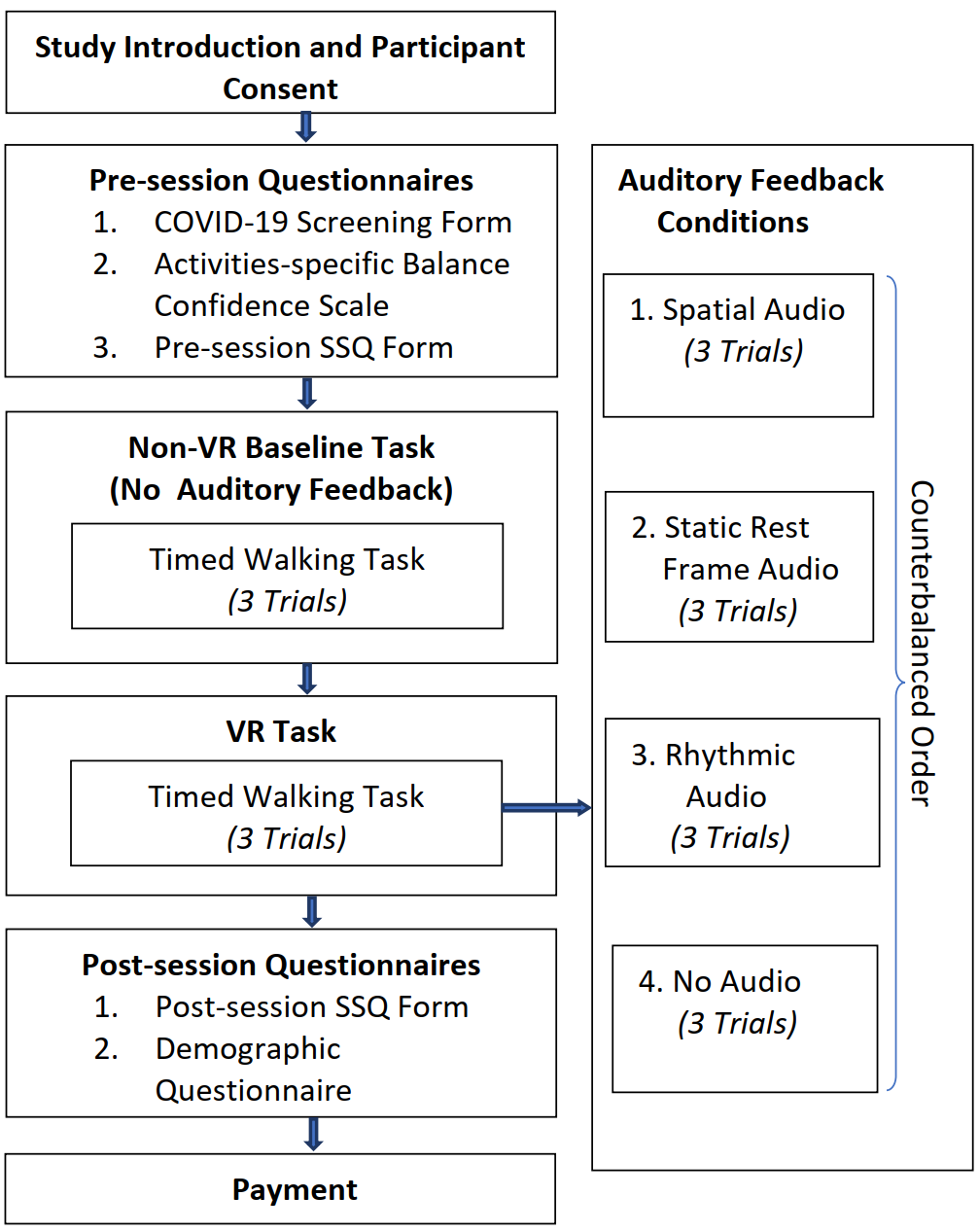}
  \caption{Study procedure}
\end{figure}

First, we sanitized all lab equipment used in the study (e.g., HMD, controllers, balance board, safety harness, and suspension system). We recorded participants' temperature upon entry to the lab and completed a COVID-19 symptom screening questionnaire form. We then informed participants of the study procedures and documented formal participant consent. Participants completed an Activities-specific Balance Confidence (ABC) \cite{powell1995activities} form and a Simulator Sickness Questionnaire (SSQ) form \cite{kennedy1993simulator} at the beginning of the study. Participants were asked to remove footwear that would interfere with the GAITRite. The Institutional Review Board (IRB) at the University of Texas at San Antonio approved our study protocols.

\subsubsection{Real World Walking}
We used the GAITRite to measure gait parameters in the study. Participants were securely fastened to the safety harness and suspension walker to prevent fall-related injuries. We instructed the participants to walk at a comfortable speed on the GAITRite. We also instructed them to complete 180 degree turns at both ends. Participants took their feet off the GAITRite while taking turns - the GAITRite software requires participants to step off between trials; moreover, the GAITRite can not technically assess turns. Participants performed three timed walking tasks \cite{steffen2002age} while we timed them using a stopwatch and collected their gait data with our GAITRite. Fig. 3 (left) shows an example of a participant’s timed walking task in the real-world environment in our study.

  

\begin{figure}[ht!]
    \centering
  \includegraphics[width=0.47\textwidth,height= 6cm]{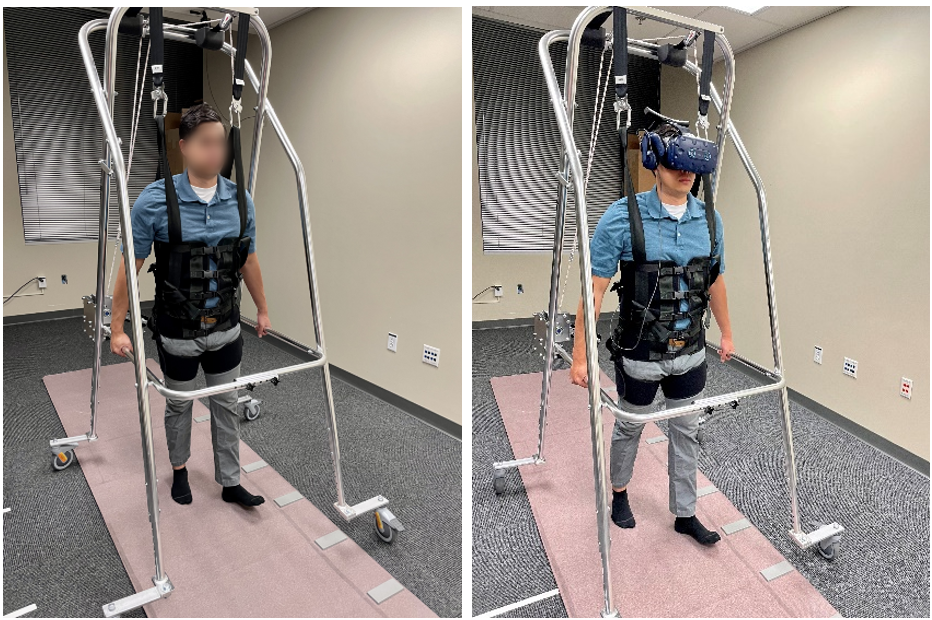}
  \caption{Participants were supported by a harness while they performed the timed walking task using the GAITRite system in real world (left, face blurred) and virtual environment (right)}
\end{figure}

\subsubsection{Virtual Environment Walking}
This was the replication of the task in the real-world environment, except it was performed in a VE with various auditory feedback conditions. We used the same harness and suspension system as the real-world environment walking to prevent sudden falls. Participants were told to walk on the virtual GAITRite overlaid on top of a real GAITRite. They used an HMD to observe the VE and the integrated HMD's headphones to hear the auditory feedback. Participants' performed three timed walking tasks in VR for each auditory condition (e.g., spatial, static rest frame, and rhythmic) and a no-audio in VR condition. The four conditions were applied in counterbalanced order for all participants. Fig. 3 (right) shows an example of a participant’s  timed walking task in the virtual environment in our study.

\subsubsection{Post-Study Questionnaires}
Participants completed a post-study SSQ form and a demographic questionnaire. Finally, each participant received compensation of \$30/Hr and a parking validation ticket at the end of the study.

\section{Metrics}

\subsection{Gait Metrics}
We investigated the following gait metrics in our study.\\
\textit{- Walking Velocity}: The distance traveled (cm) divided by ambulation time (sec).\\
\textit{- Cadence}: The number of footsteps per minute.\\
\textit{ - Step Time (Left/Right)}: The time (sec) between the initial contact points of the opposite foot.\\ 
\textit{- Step Length (Left/Right)}: The distance (cm) between heel centers of two consecutive steps of opposing feet.\\
\textit{- Cycle Time (Left/Right)}: The time (sec) between the initial contact points of two consecutive steps of the same foot.\\
\textit{- Stride Length (Left/Right)}: The distance (cm) between the steps of the same foot.\\
\textit{- Swing Time (Left/Right)}: The time (sec) between the final contact point of a foot and the initial contact point of the same foot.\\
\textit{- Stance Time (Left/Right)}: The time (sec) between the initial contact point and the final contact point of the same footstep.\\
\textit{- Single Support Time (Left/Right)}: This is the time (sec) between the current footfall's last contact and the first contact of the next footfall of the same foot. The single support time is equal to the opposing foot's swing time.\\
\textit{- Double Support Time (Left/Right)}: The time (sec) that both feet are in contact with the ground.\\
\textit{- Base of Support (Left/Right)}: The width between one foot and the line of progression of the opposite footstep. The line of progression is a line that connects the heels of two footsteps of the same foot.\\
\textit{- Toe-In/Toe-Out (Left/Right)}: The angle (degrees) between the line of progression and the center-line of a footprint. Toe-in indicates the center-line of the footprint is inside the line of progression. Toe-out indicates the center-line of the footprint is outside the line of progression.\\
More information on gait parameters can be found in the GAITRite manual \cite{WinNT3}.

\subsection{Activities-specific Balance Confidence (ABC) Scale}
The Activities-specific Balance Confidence (ABC) Scale is an outcome measure questionnaire used to assess participant balance, mobility, and physical functioning. The questionnaire uses 16 items to measure an individual’s confidence while performing everyday activities without losing balance \cite{powell1995activities}. Participants are asked to rate their confidence in each specific activity on a scale of 0\% (not confident) to 100\% (most confident). The ABC Scale score is calculated by the sum of the ratings (0-1600), divided by 16. A low level of functioning is indicated by a total ABC score below 50. A moderate level of functioning is indicated by a total ABC score between 50-80, and a high level of functioning is indicated by a total score above 80.

\subsection{Simulator Sickness Questionnaire}
The Simulator Sickness Questionnaire (SSQ) is used to measure the severity of cybersickness produced by exposure to virtual environments. The SSQ assesses participant physiological discomfort due to cybersickness using 16 symptoms in three different categories \cite{kennedy1993simulator}. The categories include nausea, oculomotor disturbance, and disorientation.

\section{Statistical Analysis}
 A Shapiro-Wilk test was applied for testing data normality for each gait parameter separately. Results indicated the normal distribution of data ( \textit{p} $>$ .05) for all gait parameters for both groups of participants. To discover any significant difference among study conditions, we used a 2$\times$5 mixed-model ANOVA with Bonferroni correction where we had one between-subject factor with two levels (participants with MI and participants without MI) and one within-subject factor with five levels (five study conditions: baseline, spatial, static, rhythmic, and no audio). When we found a significant difference, we used post hoc two-tailed paired sample t-tests to obtain the difference between two particular study conditions for within-subject comparisons and to investigate hypotheses H1, H2, and H3. To investigate hypothesis H4, we also used post hoc two-tailed t-tests between the two groups for between-subject comparisons. To assess the difference in physical ability, we used post hoc two-tailed t-tests between the ABC scores of both groups of participants. We also performed two-tailed paired sample t-tests between pre-session SSQ score and post-session SSQ score for both groups of participants separately to analyze cybersickness. Bonferroni correction was used for all tests that included multiple comparisons.

\section{Results}
Among the twelve investigated gait parameters, we found significant improvement in seven gait parameters (walking velocity, cadence, step length, stride length, step time, cycle time, and swing time) for different audio conditions and the improvement of the other five gait parameters were not significant for both groups of participants. Gait improvement also differed significantly depending on the auditory feedback conditions. We computed the gait parameters from the beginning to the end of the trials (not any specific portion of the trials). Also, we analyzed both left and right leg data. However, there was no significant difference between left and right leg data. Therefore, we reported the averaged data of the left and right leg for all gait parameters for simplicity.

We found a significant difference in walking velocity after conducting the mixed-model ANOVA test for all participants, \textit{F}(1,123) = 71.6, \textit{p} $<$ .001; and effect size, $\eta^{2}$ = 0.09. We also found a significant difference (\textit{p} $<$ .001) in cadence, step length, stride length, step time, cycle time, swing time. Next, we conducted the following post-hoc two-tailed t-tests for within-group and between-group comparisons to find differences between particular study conditions. 

\subsection{Within-Group Comparisons}

\subsubsection{Non-VR Baseline vs. VR Baseline}
For participants with MI, there was a significant decrease in walking velocity in VR baseline without audio condition (Mean, \textit{M} = 60.09, Standard Deviation, \textit{SD} = 18.97) as compared to non-VR baseline without audio condition (\textit{M} = 62.49, \textit{SD} = 18.81); \textit{t}(17) = 3.94, \textit{p} $<$ .001; and effect size, Cohen's \textit{d} = 0.12. 
For participants without MI, there was also a significant decrease in walking velocity in VR baseline without audio condition (\textit{M} = 78.79, \textit{SD} = 16.31), as compared to the non-VR baseline without audio condition (\textit{M} = 80.96, \textit{SD} = 14.73); \textit{t}(20) = 0.64, \textit{p} $<$ .001, \textit{d} = 0.13.
For both groups of participants, we also observed a significant decrease (\textit{p} $<$ .001) in cadence, step length, and stride length for VR baseline without audio condition whereas step time, cycle time, and swing time were significantly increased (\textit{p} $<$ .001) in VR baseline without audio condition as compared to non-VR baseline without audio condition. This result indicated gait disturbance in VR environments for both participants with and without MI.
 
\subsubsection{Spatial Audio vs. VR Baseline}
For participants with MI, experimental results revealed that walking velocity increased significantly more in the spatial audio condition (\textit{M} = 75.16, \textit{SD} = 21.3) as compared to VR baseline without audio condition (\textit{M} = 60.09, \textit{SD} = 18.97); \textit{t}(17) = 7.33, \textit{p} $<$ .001, \textit{d} = 0.75. 
For participants without MI, walking velocity increased significantly more in the spatial audio condition (\textit{M} = 90.35, \textit{SD} = 15.11) as compared to VR baseline without audio condition (\textit{M} = 78.79, \textit{SD} = 16.31); \textit{t}(20) = 4.72, \textit{p} $<$ .001, \textit{d} = 0.74.
For both participants with and without MI, we observed a significant increase (\textit{p} $<$ .001) in cadence, step length, and stride length for spatial audio condition as compared to VR baseline without audio condition. Also, there was a significant decrease in step time, cycle time, and swing time (\textit{p} $<$ .001) in spatial audio condition as compared to VR baseline without audio condition for both groups. This result substantiated that spatial audio improved gait parameters than the VR baseline without audio condition for both group of participants.

\begin{figure}[ht!]
    \centering
  \includegraphics[width=0.49\textwidth,height=7 cm]{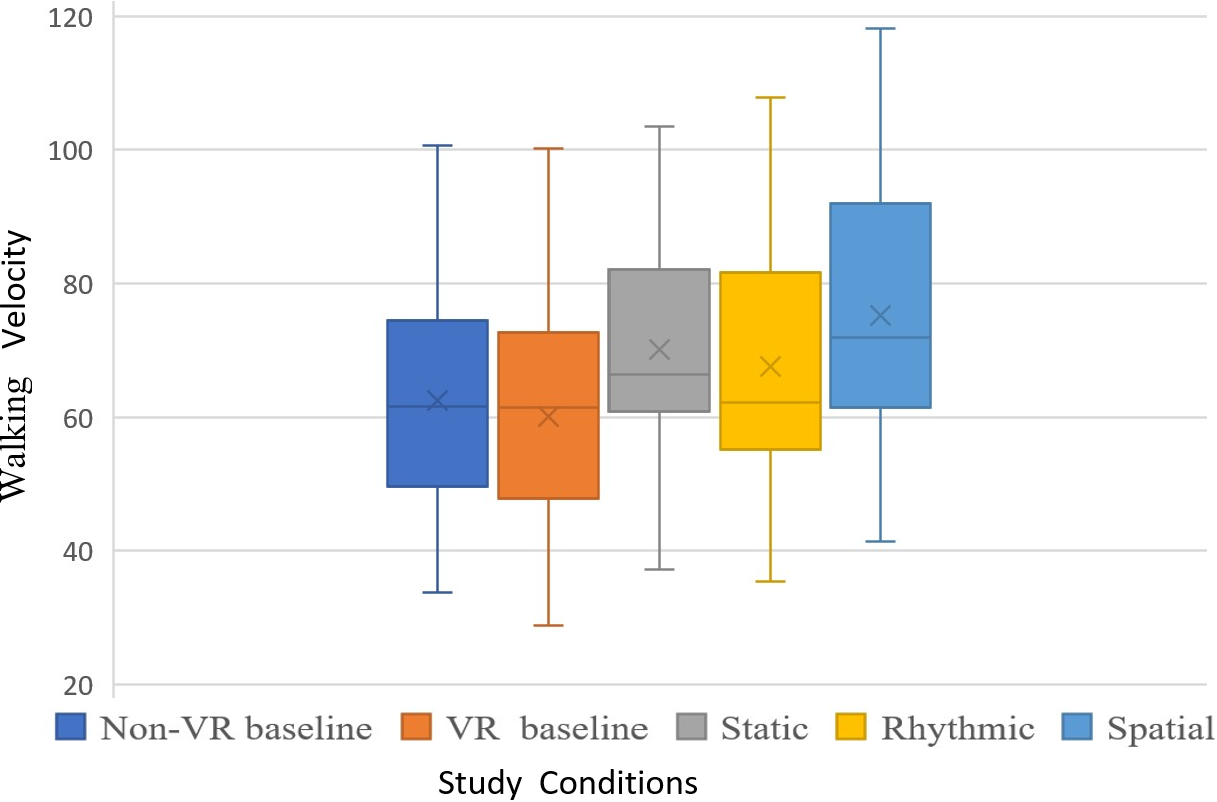}
   \caption{Walking velocity comparison between study conditions for participants with MI.}
\end{figure}

\begin{table}[ht]
\caption{Gait parameters in five conditions for participants with MI}
    \label{tab:my_label}
\begin{tabular}{|l|l|l|l|l|l|}
\hline
\begin{tabular}[c]{@{}l@{}}Gait \\ Metrics\end{tabular}           & \begin{tabular}[c]{@{}l@{}}Non-VR \\ baseline\\ \\ Mean\\ (SD)\end{tabular} & \begin{tabular}[c]{@{}l@{}}VR\\ base-\\ line\\ \\ Mean\\ (SD)\end{tabular} & \begin{tabular}[c]{@{}l@{}}Spatial\\ \\ Mean\\ (SD)\end{tabular} & \begin{tabular}[c]{@{}l@{}}Rhy-\\ thmic\\ \\ Mean\\ (SD)\end{tabular} & \begin{tabular}[c]{@{}l@{}}Static\\ Rest\\ Frame\\ \\ Mean\\ (SD)\end{tabular} \\ \hline
Cadence                                                           & \begin{tabular}[c]{@{}l@{}}74.91\\ (16.17)\end{tabular}                     & \begin{tabular}[c]{@{}l@{}}66.25\\ (19.6)\end{tabular}                     & \begin{tabular}[c]{@{}l@{}}96.99\\ (17.19)\end{tabular}          & \begin{tabular}[c]{@{}l@{}}86.43\\ (17.81)\end{tabular}           & \begin{tabular}[c]{@{}l@{}}89.23\\ (16.68)\end{tabular}                        \\ \hline
\begin{tabular}[c]{@{}l@{}}Step\\ Length\end{tabular}      & \begin{tabular}[c]{@{}l@{}}39.03\\ (7.5)\end{tabular}                      & \begin{tabular}[c]{@{}l@{}}33.53\\ (6.46)\end{tabular}                     & \begin{tabular}[c]{@{}l@{}}53.41\\ (6.99)\end{tabular}            & \begin{tabular}[c]{@{}l@{}}46.04\\ (7.19)\end{tabular}            & \begin{tabular}[c]{@{}l@{}}46.24\\ (6.69)\end{tabular}                          \\ \hline
\begin{tabular}[c]{@{}l@{}}Stride\\ Length\end{tabular}    & \begin{tabular}[c]{@{}l@{}}83.31\\ (14.92)\end{tabular}                     & \begin{tabular}[c]{@{}l@{}}77.9\\ (12.07)\end{tabular}                    & \begin{tabular}[c]{@{}l@{}}101.60\\ (14.00)\end{tabular}         & \begin{tabular}[c]{@{}l@{}}92.35\\ (14.15)\end{tabular}           & \begin{tabular}[c]{@{}l@{}}92.99\\ (12.83)\end{tabular}                        \\ \hline
\begin{tabular}[c]{@{}l@{}}Step \\ Time\end{tabular}       & \begin{tabular}[c]{@{}l@{}}0.85\\ (0.38)\end{tabular}                        & \begin{tabular}[c]{@{}l@{}}0.93\\ (0.22)\end{tabular}                      & \begin{tabular}[c]{@{}l@{}}0.52\\ (0.13)\end{tabular}            & \begin{tabular}[c]{@{}l@{}}0.72\\ (0.18)\end{tabular}             & \begin{tabular}[c]{@{}l@{}}0.72\\ (0.2)\end{tabular}                           \\ \hline
\begin{tabular}[c]{@{}l@{}}Cycle\\ Time\end{tabular}       & \begin{tabular}[c]{@{}l@{}}1.58\\ (0.33)\end{tabular}                       & \begin{tabular}[c]{@{}l@{}}1.89\\ (0.47)\end{tabular}                      & \begin{tabular}[c]{@{}l@{}}1.04\\ (0.26)\end{tabular}            & \begin{tabular}[c]{@{}l@{}}1.44\\ (0.33)\end{tabular}             & \begin{tabular}[c]{@{}l@{}}1.39\\ (0.30)\end{tabular}                          \\ \hline
\begin{tabular}[c]{@{}l@{}}Swing \\ Time\end{tabular}      & \begin{tabular}[c]{@{}l@{}}0.49\\ (0.06)\end{tabular}                       & \begin{tabular}[c]{@{}l@{}}0.56\\ (0.08)\end{tabular}                       & \begin{tabular}[c]{@{}l@{}}0.33\\ (0.06)\end{tabular}            & \begin{tabular}[c]{@{}l@{}}0.45\\ (0.07)\end{tabular}             & \begin{tabular}[c]{@{}l@{}}0.43\\ (0.06)\end{tabular}                          \\ \hline
\end{tabular}
\end{table}

\subsubsection{Spatial Audio vs. Static Rest Frame Audio}
For participants with MI, spatial audio condition increased walking velocity (\textit{M} = 75.16, \textit{SD} = 21.3) comparative to static rest frame audio condition (\textit{M} = 70.07, \textit{SD} = 18.5); \textit{t}(17) = 2.93, \textit{p} $<$ .001, \textit{d} = 0.26. 
For participants without MI, walking velocity increased significantly more in spatial audio condition (\textit{M} = 90.35, \textit{SD} = 15.11) as compared to static rest frame audio condition (\textit{M} = 84.76, \textit{SD} = 13.78); \textit{t}(20) = 3.61, \textit{p} $<$ .001, \textit{d} = 0.39.
For both participants with and without MI, we found a significant increase (\textit{p} $<$ .001) in cadence, step length, and stride length for spatial audio condition compared to static rest frame audio condition. Also, step time, cycle time,and swing time were significantly decreased (\textit{p} $<$ .001) in the spatial audio condition as compared to the static rest frame audio condition for both groups. Thus, spatial audio condition had better performance concerning gait than the static rest frame audio for participants with and without MI. 

\subsubsection{Spatial Audio vs. Rhythmic Audio}
For participants with MI, walking velocity increased significantly more in spatial audio condition (\textit{M} = 75.16, \textit{SD} = 21.3) as compared to rhythmic audio condition (\textit{M} = 67.5, \textit{SD} = 20.63); \textit{t}(17) = 4.9, \textit{p} $<$ .001, \textit{d} = 0.37. 
For participants without MI, walking velocity increased significantly more in the spatial audio condition (\textit{M} = 90.35, \textit{SD} = 15.11) as compared to the rhythmic audio condition (\textit{M} = 82.66, \textit{SD} = 15.69); \textit{t}(20) = 3.29, \textit{p} $<$ .001, \textit{d} = 0.5.
For both group of participants, there was a significant increase (\textit{p} $<$ .001) in cadence, step length, and stride length for spatial audio condition as compared to the rhythmic audio condition. However, step time, cycle time, and swing time were significantly decreased (\textit{p} $<$ .001) in spatial audio condition than rhythmic audio condition for both groups. These results suggest that spatial audio may be more effective than rhythmic audio for gait performance.

\begin{table}[ht]
\caption{Gait parameters in five conditions for participants without MI}
    \label{tab:my_label}

\begin{tabular}{|l|l|l|l|l|l|}
\hline
\begin{tabular}[c]{@{}l@{}}Gait \\ Metrics\end{tabular}         & \begin{tabular}[c]{@{}l@{}}Non-VR \\ baseline\\ \\ Mean\\ (SD)\end{tabular} & \begin{tabular}[c]{@{}l@{}}VR\\ base-\\ line\\ \\ Mean\\ (SD)\end{tabular} & \begin{tabular}[c]{@{}l@{}}Spatial\\ \\ Mean\\ (SD)\end{tabular} & \begin{tabular}[c]{@{}l@{}}Rhy-\\ thmic\\ \\ Mean\\ (SD)\end{tabular} & \begin{tabular}[c]{@{}l@{}}Static\\ Rest\\ Frame\\ \\ Mean\\ (SD)\end{tabular} \\ \hline
Cadence                                                         & \begin{tabular}[c]{@{}l@{}}90.45\\ (11.39)\end{tabular}                     & \begin{tabular}[c]{@{}l@{}}84.86\\ (13.49)\end{tabular}                    & \begin{tabular}[c]{@{}l@{}}102.79\\ (13.72)\end{tabular}         & \begin{tabular}[c]{@{}l@{}}93.42\\ (11.15)\end{tabular}               & \begin{tabular}[c]{@{}l@{}}93.04\\ (11.8)\end{tabular}                         \\ \hline
\begin{tabular}[c]{@{}l@{}}Step\\ Length\end{tabular}    & \begin{tabular}[c]{@{}l@{}}48.55\\ (6.33)\end{tabular}                      & \begin{tabular}[c]{@{}l@{}}39.46\\ (7.13)\end{tabular}                     & \begin{tabular}[c]{@{}l@{}}68.96\\ (6.3)\end{tabular}            & \begin{tabular}[c]{@{}l@{}}59.48\\ (7.71)\end{tabular}                & \begin{tabular}[c]{@{}l@{}}59.26\\ (6.98)\end{tabular}                         \\ \hline
\begin{tabular}[c]{@{}l@{}}Stride\\ Length\end{tabular}  & \begin{tabular}[c]{@{}l@{}}88.69\\ (12.67)\end{tabular}                     & \begin{tabular}[c]{@{}l@{}}83.98\\ (14.76)\end{tabular}                    & \begin{tabular}[c]{@{}l@{}}103.38\\ (12.68)\end{tabular}         & \begin{tabular}[c]{@{}l@{}}99.4\\ (14.78)\end{tabular}               & \begin{tabular}[c]{@{}l@{}}99.1\\ (13.76)\end{tabular}                         \\ \hline
\begin{tabular}[c]{@{}l@{}}Step \\ Time\end{tabular}     & \begin{tabular}[c]{@{}l@{}}0.66\\ (0.08)\end{tabular}                       & \begin{tabular}[c]{@{}l@{}}0.68\\ (0.08)\end{tabular}                      & \begin{tabular}[c]{@{}l@{}}0.55\\ (0.1)\end{tabular}             & \begin{tabular}[c]{@{}l@{}}0.63\\ (0.07)\end{tabular}                 & \begin{tabular}[c]{@{}l@{}}0.64\\ (0.08)\end{tabular}                          \\ \hline
\begin{tabular}[c]{@{}l@{}}Cycle\\ Time\end{tabular}     & \begin{tabular}[c]{@{}l@{}}1.2\\ (0.24)\end{tabular}                       & \begin{tabular}[c]{@{}l@{}}1.45\\ (0.23)\end{tabular}                       & \begin{tabular}[c]{@{}l@{}}0.98\\ (0.26)\end{tabular}            & \begin{tabular}[c]{@{}l@{}}1.16\\ (0.22)\end{tabular}                  & \begin{tabular}[c]{@{}l@{}}1.16\\ (0.23)\end{tabular}                           \\ \hline
\begin{tabular}[c]{@{}l@{}}Swing \\ Time\end{tabular}    & \begin{tabular}[c]{@{}l@{}}0.47\\ (0.05)\end{tabular}                       & \begin{tabular}[c]{@{}l@{}}0.52\\ (0.06)\end{tabular}                      & \begin{tabular}[c]{@{}l@{}}0.37\\ (0.07)\end{tabular}            & \begin{tabular}[c]{@{}l@{}}0.42\\ (0.05)\end{tabular}                 & \begin{tabular}[c]{@{}l@{}}0.42\\ (0.06)\end{tabular}                          \\ \hline
\end{tabular}
\end{table}

\subsubsection{Static Rest Frame Audio vs. VR Baseline}
For participants with MI, we observed a significant increase in walking velocity in static rest frame audio condition (\textit{M} = 70.07, \textit{SD} = 18.5) comparative to VR baseline without audio condition (\textit{M} = 60.09, \textit{SD} = 18.97); \textit{t}(17) = 8.2, \textit{p} $<$ .001, \textit{d} = 0.57. 
For participants without MI, there was a significant increase in walking velocity in static rest frame audio condition (\textit{M} = 84.76, \textit{SD} = 13.78) as compared to VR baseline without audio condition (\textit{M} = 78.79, \textit{SD} = 16.31); \textit{t}(20) = 3.89, \textit{p} $<$ .001, \textit{d} = 0.4.
For both participants with and without MI, there was a significant increase (\textit{p} $<$ .001) in cadence, step length, and stride length as compared to VR baseline without audio condition. However, step time, cycle time, and swing time were significantly decreased (\textit{p} $<$ .001) in static rest frame audio condition as compared to VR baseline without audio condition for both groups. As a result, static rest frame audio provided better performance than VR baseline without no audio condition in reference to gait behavior.

\subsubsection{Static Rest Frame Audio vs. Rhythmic Audio}
For participants with MI, there was no significant difference in walking velocity between static rest frame audio condition (\textit{M} = 70.07, \textit{SD} = 18.5) and rhythmic audio condition (\textit{M} = 67.5, \textit{SD} = 20.63); \textit{t}(17) = 1.89, \textit{p} $=$ .06, \textit{d} = 0.13 after post-hoc two-tailed paired t-test. 
For participants without MI, we did not observe a significant difference in walking velocity between static rest frame audio condition (\textit{M} = 84.76, \textit{SD} = 13.78) and rhythmic audio condition (\textit{M} = 82.66, \textit{SD} = 15.69); \textit{t}(20) = 1.11, \textit{p} $=$ .138, \textit{d} = 0.14. 
Similarly for both participants with and without MI, we did not observe any significant differences for other gait parameters between static rest frame audio condition and rhythmic audio condition. Hence, it was inconclusive which audio can be preferred between rhythmic and static rest frame for increasing gait performance.

\subsubsection{Rhythmic Audio vs. VR Baseline}
For participants with MI, tests revealed that walking velocity increased significantly more in rhythmic audio condition (\textit{M} = 67.5, \textit{SD} = 20.63) as compared to VR baseline without audio condition (\textit{M} = 60.09, \textit{SD} = 18.97); \textit{t}(17) = 5.63, \textit{p} $<$ .001, \textit{d} = 0.37. 
For participants without MI, results indicated a significant difference in walking velocity between rhythmic audio condition (\textit{M} = 82.66, \textit{SD} = 15.69) and VR baseline without audio condition (\textit{M} = 78.79, \textit{SD} = 16.31); \textit{t}(20) = 2.01, \textit{p} $<$ .001, \textit{d} = 0.24.
For both groups, cadence, step length, and stride length significantly increased (\textit{p} $<$ .001) in rhythmic audio condition as compared to VR baseline without audio condition. However, step time, cycle time, and swing time were significantly decreased (\textit{p} $<$ .001) in rhythmic audio condition as compared to VR baseline without audio condition for both groups. Thus, rhythmic auditory condition surpassed VR baseline condition for gait improvement.

The comparisons of walking velocity between five different study conditions have been shown in Figure 4 (participants with MI) and Figure 5 (participants without MI). The other gait parameters which resulted in significant improvement have been shown in Table 2 (participants with MI) and Table 3 (Participants without MI) with their respective mean and standard deviation (SD) in the five study conditions. The comparisons of effect size (Cohen's \textit{d}) for walking velocity between different study conditions have been shown in Figure 6. 

\begin{figure}[ht!]
    \centering
  \includegraphics[width=0.48\textwidth,height=7cm]{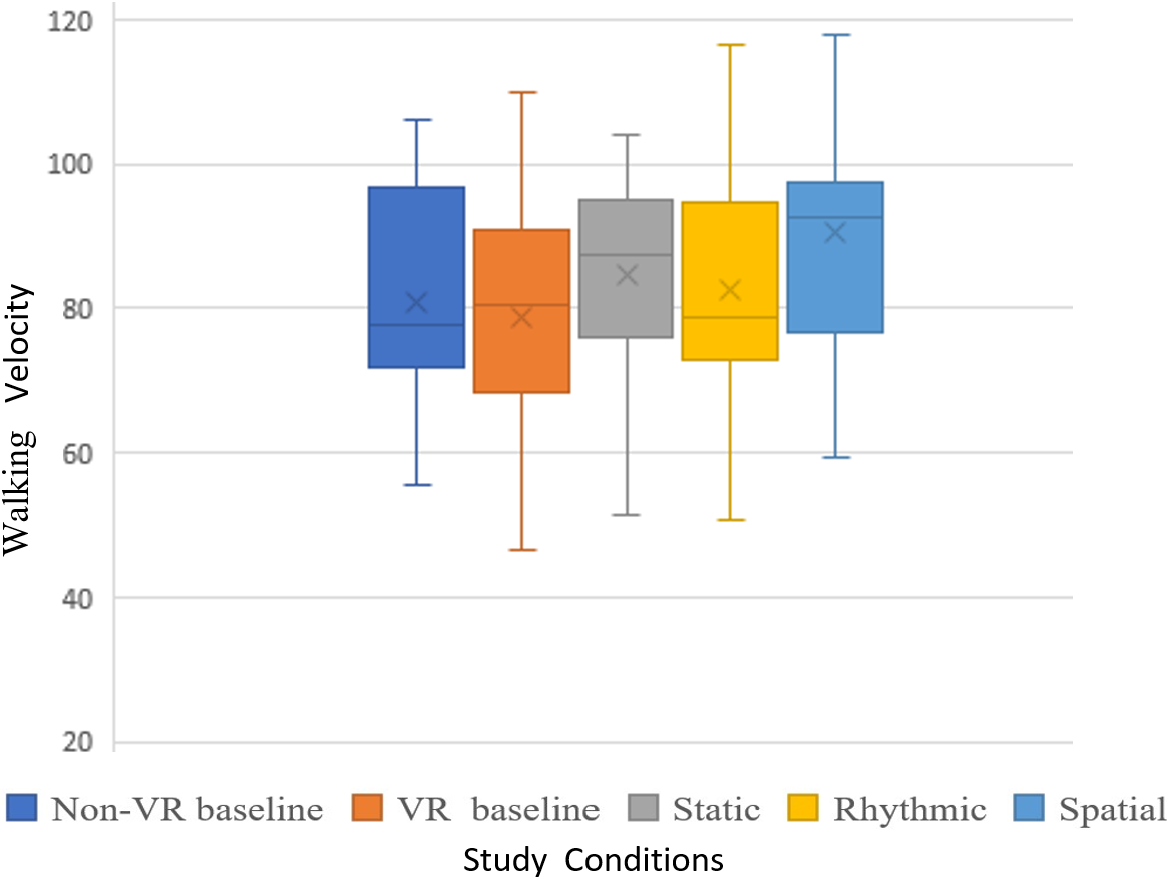}
  \caption{Walking velocity comparison between study conditions for participants without MI.}
\end{figure}

\begin{figure}[ht!]
    \centering
  \includegraphics[width=0.47\textwidth,height= 6cm]{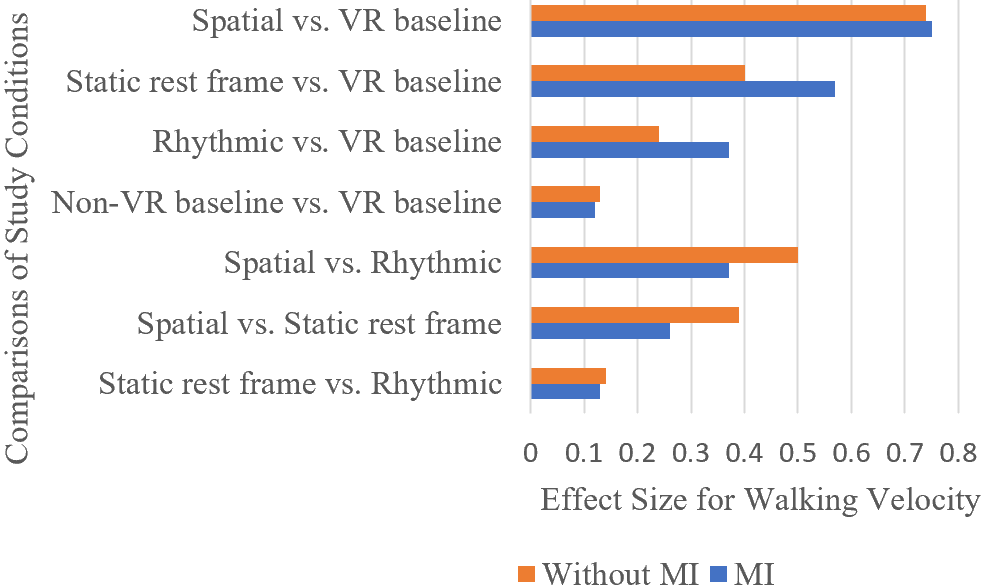}
   \caption{Comparisons of effect size for walking velocity between study conditions for participants with and without MI.}
\end{figure}

\subsection{Between-Group Comparisons}

We found a significant decrease in walking velocity for participants with MI compared to participants without MI for non-VR baseline condition; \textit{t}(37) = 3.37, \textit{p} $=$ .002; \textit{d} = 1.1 and for VR baseline condition; \textit{t}(37) = 3.27, \textit{p} $=$ .003; \textit{d} = 1.06.
Results also revealed a significant decrease in walking velocity for participants with MI as compared to participants without MI for all VR-based auditory feedback conditions: spatial audio (\textit{t}(37) = 2.53, \textit{p} $=$ .02; \textit{d} = 0.83), static rest frame audio (\textit{t}(37) = 2.77, \textit{p} $=$ .009; \textit{d} = 0.91), and for rhythmic audio (\textit{t}(37) = 2.55, \textit{p} $=$ .01; \textit{d} = 0.84).
For all conditions (non-VR baseline, VR baseline, spatial audio, static rest frame audio, and rhythmic audio), we also obtained a significant decrease (\textit{p} $<$ .05) in cadence, step length, and stride length for participants with MI than participants without MI.

\subsection{Activities-specific Balance Confidence (ABC) Scale}
We conducted a two-tailed t-test based on responses from the Activities-specific Balance Confidence (ABC) Scale between participants with MI ($M$ = 70.83, $SD$ = 24.83) and participants without MI ($M$ = 91.76, $SD$ = 13.71), \textit{t}(37) = 3.38, \textit{p} $<$ .001, \textit{d} = 1.04. The calculated mean ABC Scale score for participants with MI was 70.83\%, which indicated a moderate level of functioning. However, the calculated mean ABC score was 91.76\% for participants without MI. This indicated a high level of functioning.  These scores represented a significant difference in physical functioning between participants with and without MI. 

\subsection{Simulator Sickness Questionnaire}
We conducted a two-tailed t-test between pre-study SSQ scores and post-study SSQ scores for both participant groups. We did not observe a significant increase in SSQ scores for both participants with and without MI. We obtained \textit{t}(17) = 1.71, \textit{p} = .07, \textit{d} = 0.2 for participants with MI and \textit{t}(20) = 1.72, \textit{p} = .06, \textit{d} = 0.1 for participants without MI.


\section{Discussion}
\subsection{Gait disturbance in VR Without Audio}
Based on the results from Mixed ANOVA and post hoc t-tests for both groups of participants, we found that participants' walking velocity, step length, stride length, cadence, step time, cycle time, and swing time were significantly affected in VR-baseline without audio condition compared to the non-VR baseline without audio condition. Therefore, we noticed that gait disturbance happened in VR conditions for all participants when there was no auditory feedback, which supported our hypothesis H1. Prior works also reported that VR might cause postural instability, which could lead to gait disturbance \cite{hollman2007does,riem2020effect,sondell2005altered}.

\subsection{ Gait Improvement in VR-based Auditory Conditions }
In all VR-based auditory conditions, results showed that cadence, step length, stride length increased whereas step time, cycle time, swing time decreased. That is, participants had more steps and bigger steps in a shorter amount of time in VR-based auditory conditions as compared to the VR no-audio baseline. Thus, walking velocity was increased in all VR-based auditory conditions (Figure 4 and Figure 5) significantly more (\textit{p} $<$ .001) than VR baseline condition for both participants with and without MI, which validated our hypothesis H2. Also, the value of effect size (Cohen's \textit{d}) = 0.5 indicated that Spatial audio had a medium effect on both groups of participants. Static rest frame audio had a medium effect on participants with MI but a small effect on participants without MI. Rhythmic audio had a small effect on both groups of participants. Because of the auditory feedback effects, the gait parameters improved. Previous findings substantiated these results where they reported auditory white noise  \cite{sacco2018effects,zhou2021effects,ross2015auditory,harry2005balancing}, spatial \cite{stevens2016auditory,gandemer2017spatial}, static rest frame \cite{ross2016auditory}, CoP \cite{hasegawa2017learning}, and rhythmic audio \cite{ghai2018effect} improved gait and reduced postural sway in the real-world environment. However, most of the previous works were performed in non-VR settings while we investigated the effect of auditory feedback in VR. 

Among the VR-based auditory conditions, spatial audio outperformed (\textit{p} $<$ .001) other conditions, which supported our hypothesis H3. Also, spatial audio had a greater effect size compared to other auditory conditions (Fig. 6). This was also mentioned in previous studies where spatial audio was reported to be effective in improving gait and postural stability because it offered better fidelity \cite{chong2020audio, pinkl2020spatialized} and immersion \cite{wenzel2017perception,naef2002spatialized}. However, these studies only investigated spatial audio, whereas we compared three different kinds of auditory feedback in VR in this study.

\subsection{Gait Similarities and Dissimilarities Between Participants With and Without MI}
We found significant differences in walking velocity, cadence, step length, stride length between the participants with and without MI for all five study conditions. However, we did not find any significant difference for other gait parameters between participants with and without MI. Therefore, we found that few gait parameters (e.g., velocity, cadence, step length, stride length) were affected differently for participants with MI and participants without MI, whereas other gait parameters were affected in a similar way for both groups of participants, which supported our hypothesis H4. These results partially matched with previous work of Guo et al. \cite{guo2015mobility} where they investigated the gait parameters of participants with and without MI in a VE. They found significant differences in walking velocity, step length, and stride length between participants with MI and participants without MI, whereas there was no significant differences in other gait parameters of the two groups of participants.  

To figure out how groups differed in gait improvement from the audio conditions, we first subtracted baseline data from all conditions. Then, ANOVA and post hoc two-tailed t-tests between two different groups revealed that gait improvement for participants with MI was significantly (\textit{p} $<$ .001) more than the participants without MI. Effect size, Cohen's \textit{d} = 0.9, also indicated larger effect for participants with MI. We hypothesized that as the participants with MI had less gait functionality, there might have a chance for more improvement than the participants without MI.


\subsection{Cybersickness}
Previous research had observed that VR users exposed to virtual environments for more than 10 minutes could begin to experience the onset of cybersickness \cite{chang2020virtual,kim2021clinical}. Our study required participants to wear the HMD for around 45 minutes under several conditions, which increased the chance of developing cybersickness symptoms. However, we designed the virtual environment with no illusory self-motion to minimize the possibility of participants developing cybersickness \cite{mccauley1992cybersickness}.  We learned from the post-study verbal conversation with the participants that a few experienced mild cybersickness after the study, which only raised their post SSQ score slightly. Moreover, there was no significant difference between pre-SSQ and post-SSQ scores. This suggests that cybersickness was negligible and did not alter the participant's gait performance. 


\section{Limitations}

Both participant groups used a suspension walking system consisting of a body harness and thigh cuffs for the duration of the timed walking task. The walking system was used in the baseline condition and all auditory feedback conditions in VR by every participant to maintain study procedure consistency. Although participants were instructed to walk at a comfortable speed, the heavy suspension walking system could have reduced participants' normal walking speed. Previous studies also reported that wearing a safety harness caused a significant decrease in walking speed \cite{decker2012wearing}. Due to this intervention, studies that do not require a safety harness may observe different outcomes. 

The duration of the study was lengthy and required participants to complete multiple time-consuming trials on the GAITRite which could sometimes produce symptoms of fatigue in participants. To mitigate the fatigue effect, participants were able to rest and remove the HMD between trials and conditions if needed. This rest and removal of the HMD may have allowed them to regain spatial orientation of the room setup and possibly skewed data results. 


In our study, we applied the four auditory feedback conditions in counterbalanced order, which reduces carryover or practice effects \cite{WinNT5, WinNT4, WinNT6}. Counterbalancing order was also reported to be effective in many prior research\cite{millner2020suicide,plechata2019age,sheskin2018thechildlab}. Alternatively, we could have applied the auditory feedback in randomized order to reduce bias. However, our study included participants with MS who are very prone to fatigue and cybersickness. Thus, we were more concerned about the carryover fatigue and cybersickness effects on results. As counterbalancing also reduces fatigue and cybersickness effects\cite{WinNT6}, we preferred counterbalancing over randomization.

We applied the "rhythmic" auditory feedback in every one-second interval. However, we did not investigate this feedback condition for other time intervals (e.g., two-second). Therefore, studies that would apply "rhythmic" auditory feedback for different time intervals might find slightly different results for this specific condition. 

For static rest frame audio condition, continuously played white noise could have had fatigue effect on participants. However, we did not measure the fatigue effect for this condition.

In our study, the non-VR baseline was always done first, which might have influenced the walking speed for this condition. However, we wanted to have enough baseline tasks before starting the VR conditions to reduce the learning effect of VR conditions.

Our research focus was to investigate solutions to solve the gait disturbance issues in VR. So, we did not consider non-VR audio conditions. Also, adding the three auditory feedback conditions in non-VR would result in three additional study conditions, which would make the study significantly longer. As our study included participants with MI due to MS who had less physical ability and were prone to fatigue, we tried to keep the study time shorter. 

We had more female participants than males in our study. This is because we recruited from the population with MS, which is statistically more common in females \cite{WinNT}. Many previous studies reported no significant effect of gender on balance \cite{kahraman2018gender, faraldo2012influence,schedler2019age}. However, we plan to investigate the gender effect on balance in VR in our future work.

We measured gait performance during VR intervention. We did not measure the post-study effects on gait. Our motivation here was accessibility rather than rehabilitation, and thus we only investigated gait outside of VR as a baseline and during VR immersion.

We had five different study conditions. We performed three trials for each study condition, which resulted in fifteen trials for each participant. We collected separate data files for each trial. Therefore, we performed total of 585 trials for our 39 participants. The HMD display stopped working during four trials of four participants (MI group:3, Healthy group: 1). Restarting the "Vive wireless app" solved the issue each time. We repeated those four incomplete trials and omitted the four incomplete data files. 

We encountered challenges with participant recruitment due to the COVID-19 pandemic. Many of our participants with MI had multiple sclerosis and were therefore immunosuppressed. This placed them in the “high-risk” category for contracting COVID-19 and deterred them from joining the study. We would have been able to recruit a larger participant group and provide additional gait performance results if the study was conducted at a period outside of the pandemic.


\section{Conclusion}
We found significant evidence that spatial, rhythmic, and static rest frame auditory feedback conditions resulted in the improvement of gait performance in both participant groups while immersed in VR. Spatial audio improved gait parameters significantly more than rhythmic and static rest frame audio conditions. Also, improvements of gait parameters were significantly greater in participants with mobility impairments than the participants without mobility impairments. The results from this study will provide guidance to researchers to better understand the implications of assistive technologies based on auditory feedback for improving gait performance in HMD-based VEs. Furthermore, these results suggest that auditory feedback should be considered more in the future development of VR experiences to improve usability and accessibility, especially for persons with mobility impairments.


\acknowledgments{
The US National Science Foundation provided funding for this project (IIS 2007041). We also express our gratitude to all of our research participants.
}

\bibliographystyle{abbrv-doi}

\bibliography{template}

\begin{thebibliography}{10}

\bibitem{WinNT4}
{[Counterbalancing]} counterbalancing to reduce carryover effect.
\newblock
  \url{https://opentext.wsu.edu/carriecuttler/chapter/experimental-design/}.
\newblock Accessed: 2021-08-30.

\bibitem{WinNT6}
{[Counterbalancing]} counterbalancing to reduce fatigue and practice effect.
\newblock
  \url{https://psychology.illinoisstate.edu/jccutti/psych231/SP01/week8.html#:~:text=Advantages%3A%20The%20biggest%20advantage%20is%20that%20exposure%20to,levels%20of%20the%20IV%20Counterbalancing%20is%20not%20required}.
\newblock Accessed: 2021-08-30.

\bibitem{WinNT5}
{[Counterbalancing]} one of the best ways to avoid pitfalls of repeated measure
  designs.
\newblock
  \url{https://explorable.com/counterbalanced-measures-design#:~:text=Experiments%20conducted%20with%20a%20counterbalanced%20measures%20design%20are,subjects%20are%20exposed%20to%20all%20of%20the%20treatments.}
\newblock Accessed: 2021-08-30.

\bibitem{WinNT3}
{[GaitRite Manual]} details gait parameters for gaitrite walkway system.
\newblock
  \url{https://www.procarebv.nl/wp-content/uploads/2017/01/Technische-aspecten-GAITrite-Walkway-System.pdf}.
\newblock Accessed: 2021-08-30.

\bibitem{WinNT1}
{[n. d.]} google resonance audio.
\newblock
  \url{https://resonance-audio.github.io/resonance-audio/develop/unity/getting-started}.
\newblock Accessed: 2021-08-30.

\bibitem{WinNT}
{[n. d.]} people with disabilities in the world.
\newblock
  \url{https://www.un.org/development/desa/disabilities/resources/factsheet-on-persons-with-disabilities.html}.
\newblock Accessed: 2021-08-30.

\bibitem{agrawal2009disorders}
Y.~Agrawal, J.~P. Carey, C.~C. Della~Santina, M.~C. Schubert, and L.~B. Minor.
\newblock Disorders of balance and vestibular function in us adults: data from
  the national health and nutrition examination survey, 2001-2004.
\newblock {\em Archives of internal medicine}, 169(10):938--944, 2009.

\bibitem{baram2007auditory}
Y.~Baram and A.~Miller.
\newblock Auditory feedback control for improvement of gait in patients with
  multiple sclerosis.
\newblock {\em Journal of the neurological sciences}, 254(1-2):90--94, 2007.

\bibitem{bergeron2015use}
M.~Bergeron, C.~L. Lortie, and M.~J. Guitton.
\newblock Use of virtual reality tools for vestibular disorders rehabilitation:
  a comprehensive analysis.
\newblock {\em Advances in medicine}, 2015, 2015.

\bibitem{vcakrt2010exercise}
O.~{\v{C}}akrt, M.~Chovanec, T.~Funda, P.~Kalitov{\'a}, J.~Betka,
  E.~Zv{\v{e}}{\v{r}}ina, P.~Kol{\'a}{\v{r}}, and J.~Je{\v{r}}{\'a}bek.
\newblock Exercise with visual feedback improves postural stability after
  vestibular schwannoma surgery.
\newblock {\em European archives of oto-rhino-laryngology}, 267(9):1355--1360,
  2010.

\bibitem{canessa2019comparing}
A.~Canessa, P.~Casu, F.~Solari, and M.~Chessa.
\newblock Comparing real walking in immersive virtual reality and in physical
  world using gait analysis.
\newblock In {\em VISIGRAPP (2: HUCAPP)}, pp. 121--128, 2019.

\bibitem{chang2020virtual}
E.~Chang, H.~T. Kim, and B.~Yoo.
\newblock Virtual reality sickness: a review of causes and measurements.
\newblock {\em International Journal of Human--Computer Interaction},
  36(17):1658--1682, 2020.

\bibitem{chiari2005audio}
L.~Chiari, M.~Dozza, A.~Cappello, F.~B. Horak, V.~Macellari, and D.~Giansanti.
\newblock Audio-biofeedback for balance improvement: an accelerometry-based
  system.
\newblock {\em IEEE transactions on biomedical engineering}, 52(12):2108--2111,
  2005.

\bibitem{cho2016treadmill}
C.~Cho, W.~Hwang, S.~Hwang, and Y.~Chung.
\newblock Treadmill training with virtual reality improves gait, balance, and
  muscle strength in children with cerebral palsy.
\newblock {\em The Tohoku journal of experimental medicine}, 238(3):213--218,
  2016.

\bibitem{chong2020audio}
U.~Chong and S.~Alimardanov.
\newblock Audio augmented reality using unity for marine tourism.
\newblock In {\em International Conference on Intelligent Human Computer
  Interaction}, pp. 303--311. Springer, 2020.

\bibitem{cornwell2020walking}
T.~Cornwell, J.~Woodward, M.~Wu, B.~Jackson, P.~Souza, J.~Siegel, S.~Dhar,
  K.~E. Gordon, et~al.
\newblock Walking with ears: altered auditory feedback impacts gait step length
  in older adults.
\newblock {\em Frontiers in Sports and Active Living}, 2:38, 2020.

\bibitem{de2016effect}
I.~J. De~Rooij, I.~G. Van De~Port, and J.-W.~G. Meijer.
\newblock Effect of virtual reality training on balance and gait ability in
  patients with stroke: systematic review and meta-analysis.
\newblock {\em Physical therapy}, 96(12):1905--1918, 2016.

\bibitem{decker2012wearing}
L.~M. Decker, F.~Cignetti, and N.~Stergiou.
\newblock Wearing a safety harness during treadmill walking influences lower
  extremity kinematics mainly through changes in ankle regularity and local
  stability.
\newblock {\em Journal of NeuroEngineering and Rehabilitation}, 9(1):1--8,
  2012.

\bibitem{duque2013effects}
G.~Duque, D.~Boersma, G.~Loza-Diaz, S.~Hassan, H.~Suarez, D.~Geisinger,
  P.~Suriyaarachchi, A.~Sharma, and O.~Demontiero.
\newblock Effects of balance training using a virtual-reality system in older
  fallers.
\newblock {\em Clinical interventions in aging}, 8:257, 2013.

\bibitem{epure2014effect}
P.~Epure, C.~Gheorghe, T.~Nissen, L.-O. Toader, A.~Nicolae, S.~S. Nielsen,
  D.~J.~R. Christensen, A.~L. Brooks, and E.~Petersson.
\newblock Effect of the oculus rift head mounted display on postural stability.
\newblock In {\em The 10th International Conference on Disability Virtual
  Reality \& Associated Technologies: Proceedings}, pp. 119--127. Reading
  University Press, 2014.

\bibitem{faraldo2012influence}
A.~Faraldo-Garc{\'\i}a, S.~Santos-P{\'e}rez, R.~Crujeiras-Casais,
  T.~Labella-Caballero, and A.~Soto-Varela.
\newblock Influence of age and gender in the sensory analysis of balance
  control.
\newblock {\em European Archives of Oto-Rhino-Laryngology}, 269(2):673--677,
  2012.

\bibitem{ferdous2018investigating}
S.~M.~S. Ferdous, T.~I. Chowdhury, I.~M. Arafat, and J.~Quarles.
\newblock Investigating the reason for increased postural instability in
  virtual reality for persons with balance impairments.
\newblock In {\em Proceedings of the 24th ACM Symposium on Virtual Reality
  Software and Technology}, pp. 1--7, 2018.

\bibitem{franco2012ibalance}
C.~Franco, A.~Fleury, P.-Y. Gum{\'e}ry, B.~Diot, J.~Demongeot, and
  N.~Vuillerme.
\newblock ibalance-abf: a smartphone-based audio-biofeedback balance system.
\newblock {\em IEEE transactions on biomedical engineering}, 60(1):211--215,
  2012.

\bibitem{gandemer2016sound}
L.~Gandemer, G.~Parseihian, C.~Bourdin, and R.~Kronland-Martinet.
\newblock Sound and posture: an overview of recent findings.
\newblock In {\em Computer Music and Multidisciplinary Reasearch (CMMR) 2016},
  2016.

\bibitem{gandemer2017spatial}
L.~Gandemer, G.~Parseihian, R.~Kronland-Martinet, and C.~Bourdin.
\newblock Spatial cues provided by sound improve postural stabilization:
  evidence of a spatial auditory map?
\newblock {\em Frontiers in neuroscience}, 11:357, 2017.

\bibitem{ghai2018effect}
S.~Ghai, I.~Ghai, and A.~O. Effenberg.
\newblock Effect of rhythmic auditory cueing on aging gait: a systematic review
  and meta-analysis.
\newblock {\em Aging and disease}, 9(5):901, 2018.

\bibitem{guo2013effects}
R.~Guo, G.~Samaraweera, and J.~Quarles.
\newblock The effects of ves on mobility impaired users: Presence, gait, and
  physiological response.
\newblock In {\em Proceedings of the 19th ACM Symposium on Virtual Reality
  Software and Technology}, pp. 59--68, 2013.

\bibitem{guo2015mobility}
R.~Guo, G.~Samaraweera, and J.~Quarles.
\newblock Mobility impaired users respond differently than healthy users in
  virtual environments.
\newblock {\em Computer Animation and Virtual Worlds}, 26(5):509--526, 2015.

\bibitem{harry2005balancing}
J.~D. Harry, J.~B. Niemi, A.~A. Priplata, and J.~Collins.
\newblock Balancing act [noise based sensory enhancement technology].
\newblock {\em IEEE Spectrum}, 42(4):36--41, 2005.

\bibitem{hasegawa2017learning}
N.~Hasegawa, K.~Takeda, M.~Sakuma, H.~Mani, H.~Maejima, and T.~Asaka.
\newblock Learning effects of dynamic postural control by auditory biofeedback
  versus visual biofeedback training.
\newblock {\em Gait \& posture}, 58:188--193, 2017.

\bibitem{helps2014different}
S.~K. Helps, S.~Bamford, E.~J. Sonuga-Barke, and G.~B. S{\"o}derlund.
\newblock Different effects of adding white noise on cognitive performance of
  sub-, normal and super-attentive school children.
\newblock {\em PloS one}, 9(11):e112768, 2014.

\bibitem{hollman2007does}
J.~H. Hollman, R.~H. Brey, T.~J. Bang, and K.~R. Kaufman.
\newblock Does walking in a virtual environment induce unstable gait?: An
  examination of vertical ground reaction forces.
\newblock {\em Gait \& posture}, 26(2):289--294, 2007.

\bibitem{horlings2009influence}
C.~G. Horlings, M.~G. Carpenter, U.~M. K{\"u}ng, F.~Honegger, B.~Wiederhold,
  and J.~H. Allum.
\newblock Influence of virtual reality on postural stability during movements
  of quiet stance.
\newblock {\em Neuroscience letters}, 451(3):227--231, 2009.

\bibitem{horsak2021overground}
B.~Horsak, M.~Simonlehner, L.~Sch{\"o}ffer, B.~Dumphart, A.~Jalaeefar, and
  M.~Husinsky.
\newblock Overground walking in a fully immersive virtual reality: A
  comprehensive study on the effects on full-body walking biomechanics.
\newblock {\em Frontiers in bioengineering and biotechnology}, 9, 2021.

\bibitem{janeh2019gait}
O.~Janeh, O.~Fr{\"u}ndt, B.~Sch{\"o}nwald, A.~Gulberti, C.~Buhmann, C.~Gerloff,
  F.~Steinicke, and M.~P{\"o}tter-Nerger.
\newblock Gait training in virtual reality: short-term effects of different
  virtual manipulation techniques in parkinson’s disease.
\newblock {\em Cells}, 8(5):419, 2019.

\bibitem{janeh2021review}
O.~Janeh and F.~Steinicke.
\newblock A review of the potential of virtual walking techniques for gait
  rehabilitation.
\newblock {\em Frontiers in Human Neuroscience}, 15, 2021.

\bibitem{kahraman2018gender}
B.~O. Kahraman, T.~Kahraman, O.~Kalemci, and Y.~S. Sengul.
\newblock Gender differences in postural control in people with nonspecific
  chronic low back pain.
\newblock {\em Gait \& Posture}, 64:147--151, 2018.

\bibitem{kennedy1993simulator}
R.~S. Kennedy, N.~E. Lane, K.~S. Berbaum, and M.~G. Lilienthal.
\newblock Simulator sickness questionnaire: An enhanced method for quantifying
  simulator sickness.
\newblock {\em The international journal of aviation psychology},
  3(3):203--220, 1993.

\bibitem{kim2021clinical}
H.~Kim, D.~J. Kim, W.~H. Chung, K.-A. Park, J.~D. Kim, D.~Kim, K.~Kim, and
  H.~J. Jeon.
\newblock Clinical predictors of cybersickness in virtual reality (vr) among
  highly stressed people.
\newblock {\em Scientific reports}, 11(1):1--11, 2021.

\bibitem{lott2003effect}
A.~Lott, E.~Bisson, Y.~Lajoie, J.~McComas, and H.~Sveistrup.
\newblock The effect of two types of virtual reality on voluntary center of
  pressure displacement.
\newblock {\em Cyberpsychology \& behavior}, 6(5):477--485, 2003.

\bibitem{maculewicz2015effects}
J.~Maculewicz, A.~Jylh{\"a}, S.~Serafin, and C.~Erkut.
\newblock The effects of ecological auditory feedback on rhythmic walking
  interaction.
\newblock {\em IEEE MultiMedia}, 22(1):24--31, 2015.

\bibitem{mahmud2022auditory}
M.~R. Mahmud, M.~Stewart, A.~Cordova, and J.~Quarles.
\newblock Auditory feedback for standing balance improvement in virtual
  reality.
\newblock In {\em 2022 IEEE Conference on Virtual Reality and 3D User
  Interfaces (VR)}, pp. 782--791. IEEE, 2022.

\bibitem{mahmud2022vibrotactile}
M.~R. Mahmud, M.~Stewart, A.~Cordova, and J.~Quarles.
\newblock Vibrotactile feedback to make real walking in virtual reality more
  accessible.
\newblock {\em arXiv preprint arXiv:2208.02403}, 2022.

\bibitem{martelli2019gait}
D.~Martelli, B.~Xia, A.~Prado, and S.~K. Agrawal.
\newblock Gait adaptations during overground walking and multidirectional
  oscillations of the visual field in a virtual reality headset.
\newblock {\em Gait \& posture}, 67:251--256, 2019.

\bibitem{martinez2018analysing}
A.~Martinez, A.~I. Paganelli, and A.~Raposo.
\newblock Analysing balance loss in vr interaction with hmds.
\newblock {\em Journal on Interactive Systems}, 9(2), 2018.

\bibitem{mccauley1992cybersickness}
M.~E. McCauley and T.~J. Sharkey.
\newblock Cybersickness: Perception of self-motion in virtual environments.
\newblock {\em Presence: Teleoperators \& Virtual Environments}, 1(3):311--318,
  1992.

\bibitem{meldrum2012effectiveness}
D.~Meldrum, S.~Herdman, R.~Moloney, D.~Murray, D.~Duffy, K.~Malone, H.~French,
  S.~Hone, R.~Conroy, and R.~McConn-Walsh.
\newblock Effectiveness of conventional versus virtual reality based vestibular
  rehabilitation in the treatment of dizziness, gait and balance impairment in
  adults with unilateral peripheral vestibular loss: a randomised controlled
  trial.
\newblock {\em BMC Ear, Nose and Throat Disorders}, 12(1):1--8, 2012.

\bibitem{millner2020suicide}
A.~J. Millner, M.~D. Lee, K.~Hoyt, J.~W. Buckholtz, R.~P. Auerbach, and M.~K.
  Nock.
\newblock Are suicide attempters more impulsive than suicide ideators?
\newblock {\em General hospital psychiatry}, 63:103--110, 2020.

\bibitem{murata2004effects}
A.~Murata.
\newblock Effects of duration of immersion in a virtual reality environment on
  postural stability.
\newblock {\em International Journal of Human-Computer Interaction},
  17(4):463--477, 2004.

\bibitem{naef2002spatialized}
M.~Naef, O.~Staadt, and M.~Gross.
\newblock Spatialized audio rendering for immersive virtual environments.
\newblock In {\em Proceedings of the ACM symposium on Virtual reality software
  and technology}, pp. 65--72, 2002.

\bibitem{nilsson2018natural}
N.~C. Nilsson, S.~Serafin, F.~Steinicke, and R.~Nordahl.
\newblock Natural walking in virtual reality: A review.
\newblock {\em Computers in Entertainment (CIE)}, 16(2):1--22, 2018.

\bibitem{park2015effects}
E.-C. Park, S.-G. Kim, and C.-W. Lee.
\newblock The effects of virtual reality game exercise on balance and gait of
  the elderly.
\newblock {\em Journal of physical therapy science}, 27(4):1157--1159, 2015.

\bibitem{pinkl2020spatialized}
J.~Pinkl and M.~Cohen.
\newblock Spatialized ar polyrhythmic metronome using bose frames eyewear.
\newblock 2020.

\bibitem{plechata2019age}
A.~Plechat{\'a}, V.~Sahula, D.~Fayette, and I.~Fajnerov{\'a}.
\newblock Age-related differences with immersive and non-immersive virtual
  reality in memory assessment.
\newblock {\em Frontiers in psychology}, p. 1330, 2019.

\bibitem{powell1995activities}
L.~E. Powell and A.~M. Myers.
\newblock The activities-specific balance confidence (abc) scale.
\newblock {\em The Journals of Gerontology Series A: Biological Sciences and
  Medical Sciences}, 50(1):M28--M34, 1995.

\bibitem{riem2020effect}
L.~I. Riem, B.~D. Schmit, and S.~A. Beardsley.
\newblock The effect of discrete visual perturbations on balance control during
  gait.
\newblock In {\em 2020 42nd Annual International Conference of the IEEE
  Engineering in Medicine \& Biology Society (EMBC)}, pp. 3162--3165. IEEE,
  2020.

\bibitem{robert2016effect}
M.~T. Robert, L.~Ballaz, and M.~Lemay.
\newblock The effect of viewing a virtual environment through a head-mounted
  display on balance.
\newblock {\em Gait \& posture}, 48:261--266, 2016.

\bibitem{ross2016auditory}
J.~Ross, O.~Will, Z.~McGann, and R.~Balasubramaniam.
\newblock Auditory white noise reduces age-related fluctuations in balance.
\newblock {\em Neuroscience letters}, 630:216--221, 2016.

\bibitem{ross2015auditory}
J.~M. Ross and R.~Balasubramaniam.
\newblock Auditory white noise reduces postural fluctuations even in the
  absence of vision.
\newblock {\em Experimental brain research}, 233(8):2357--2363, 2015.

\bibitem{sacco2018effects}
C.~C. Sacco, E.~M. Gaffney, and J.~C. Dean.
\newblock Effects of white noise achilles tendon vibration on quiet standing
  and active postural positioning.
\newblock {\em Journal of applied biomechanics}, 34(2):151--158, 2018.

\bibitem{samaraweera2015applying}
G.~Samaraweera, A.~Perdomo, and J.~Quarles.
\newblock Applying latency to half of a self-avatar's body to change real
  walking patterns.
\newblock In {\em 2015 IEEE Virtual Reality (VR)}, pp. 89--96. IEEE, 2015.

\bibitem{schedler2019age}
S.~Schedler, R.~Kiss, and T.~Muehlbauer.
\newblock Age and sex differences in human balance performance from 6-18 years
  of age: a systematic review and meta-analysis.
\newblock {\em PLoS one}, 14(4):e0214434, 2019.

\bibitem{sheskin2018thechildlab}
M.~Sheskin and F.~Keil.
\newblock Thechildlab. com a video chat platform for developmental research.
\newblock 2018.

\bibitem{sienko2017role}
K.~Sienko, S.~Whitney, W.~Carender, and C.~Wall~III.
\newblock The role of sensory augmentation for people with vestibular deficits:
  real-time balance aid and/or rehabilitation device?
\newblock {\em Journal of Vestibular Research}, 27(1):63--76, 2017.

\bibitem{soltani2020influence}
P.~Soltani and R.~Andrade.
\newblock The influence of virtual reality head-mounted displays on balance
  outcomes and training paradigms: A systematic review.
\newblock {\em Frontiers in sports and active living}, 2:233, 2020.

\bibitem{sondell2005altered}
B.~Sondell, L.~Nyberg, S.~Eriksson, B.~Engstr{\"o}m, A.~Backman, K.~Holmlund,
  G.~Bucht, and L.~Lundin-Olsson.
\newblock Altered walking pattern in a virtual environment.
\newblock {\em Presence}, 14(2):191--197, 2005.

\bibitem{steffen2002age}
T.~M. Steffen, T.~A. Hacker, and L.~Mollinger.
\newblock Age-and gender-related test performance in community-dwelling elderly
  people: Six-minute walk test, berg balance scale, timed up \& go test, and
  gait speeds.
\newblock {\em Physical therapy}, 82(2):128--137, 2002.

\bibitem{stevens2016auditory}
M.~N. Stevens, D.~L. Barbour, M.~P. Gronski, and T.~E. Hullar.
\newblock Auditory contributions to maintaining balance.
\newblock {\em Journal of Vestibular Research}, 26(5-6):433--438, 2016.

\bibitem{sutbeyaz2007mirror}
S.~S{\"u}tbeyaz, G.~Yavuzer, N.~Sezer, and B.~F. Koseoglu.
\newblock Mirror therapy enhances lower-extremity motor recovery and motor
  functioning after stroke: a randomized controlled trial.
\newblock {\em Archives of physical medicine and rehabilitation},
  88(5):555--559, 2007.

\bibitem{takahashi2001change}
Y.~Takahashi and A.~Murata.
\newblock Change of equilibrium under the influence of vr experience.
\newblock In {\em Proceedings 10th IEEE International Workshop on Robot and
  Human Interactive Communication. ROMAN 2001 (Cat. No. 01TH8591)}, pp.
  642--647. IEEE, 2001.

\bibitem{thikey2011need}
H.~Thikey, F.~van Wjick, M.~Grealy, and P.~Rowe.
\newblock A need for meaningful visual feedback of lower extremity function
  after stroke.
\newblock In {\em 2011 5th International Conference on Pervasive Computing
  Technologies for Healthcare (PervasiveHealth) and Workshops}, pp. 379--383.
  IEEE, 2011.

\bibitem{velazquez2010wearable}
R.~Vel{\'a}zquez.
\newblock Wearable assistive devices for the blind.
\newblock In {\em Wearable and autonomous biomedical devices and systems for
  smart environment}, pp. 331--349. Springer, 2010.

\bibitem{walker2010virtual}
M.~L. Walker, S.~I. Ringleb, G.~C. Maihafer, R.~Walker, J.~R. Crouch,
  B.~Van~Lunen, and S.~Morrison.
\newblock Virtual reality--enhanced partial body weight--supported treadmill
  training poststroke: feasibility and effectiveness in 6 subjects.
\newblock {\em Archives of physical medicine and rehabilitation},
  91(1):115--122, 2010.

\bibitem{wenzel2017perception}
E.~M. Wenzel, D.~R. Begault, and M.~Godfroy-Cooper.
\newblock Perception of spatial sound.
\newblock In {\em Immersive sound}, pp. 5--39. Routledge, 2017.

\bibitem{winter2021immersive}
C.~Winter, F.~Kern, D.~Gall, M.~E. Latoschik, P.~Pauli, and I.~K{\"a}thner.
\newblock Immersive virtual reality during gait rehabilitation increases
  walking speed and motivation: a usability evaluation with healthy
  participants and patients with multiple sclerosis and stroke.
\newblock {\em Journal of neuroengineering and rehabilitation}, 18(1):1--14,
  2021.

\bibitem{zhou2021effects}
Z.~Zhou, C.~Wu, Z.~Hu, Y.~Chai, K.~Chen, and T.~Asakawa.
\newblock Effects of white gaussian noise on dynamic balance in healthy young
  adults.
\newblock {\em Scientific reports}, 11(1):1--10, 2021.

\end{thebibliography}
\end{document}